\documentclass[aps,prb,floatfix,twocolumn]{revtex4-2}
\usepackage{amsmath, amssymb, physics, graphicx}
\usepackage[margin=0.7in]{geometry}
\usepackage{tabularx}
\usepackage[table]{xcolor}
\usepackage{array}
\usepackage{mathtools} 
\usepackage[colorlinks=true, linkcolor=blue, citecolor=red, urlcolor=blue]{hyperref}
\newcolumntype{Y}{>{\centering\arraybackslash}X}

\begin{document}
\title{A Qubit as a Bridge Between Statistical Mechanics and Quantum Dynamics}
\author{Manmeet Kaur}
\email{manmeet.kaur_phd20\string@ashoka.edu.in}
\author{Somendra M. Bhattacharjee}
\email{somendra.bhattacharjee\string@ashoka.edu.in}
\affiliation{Department of Physics, Ashoka University, Sonepat,  Haryana - 131029, India}
%\date{\today}
\begin{abstract}
This work presents a unified perspective on thermal equilibrium and
quantum dynamics by examining the simplest quantum system, a qubit, as
a minimal model. We show that both the thermal partition function
and the Loschmidt amplitude can be understood as extensions of a
single analytic function along different paths in the complex
plane. The zeros of Loschmidt amplitude encode dynamical features such
as orthogonality, rate function singularities, and quantum speed
limits, in analogy with the role of partition function zeros in
equilibrium statistical mechanics. We further establish, through the
Cauchy-Riemann equations, that the high-temperature specific heat
corresponds to early-time evolution. The discussion follows a
pedagogical progression from a single qubit to an interacting spin
chain, all with finite dimensional Hilbert spaces.

\vspace{1.5em} 
\end{abstract}
\maketitle
\section {Introduction}
The boundary between thermal physics and quantum dynamics is often
drawn sharply; the former is related to equilibrium ensembles, while
the latter describes unitary time evolution. Yet at a fundamental
level, both frameworks aim to understand how physical systems access
and evolve through their available states. In this work, we
investigate how a single quantum system, \textit{a qubit},
representing a two-level system~\cite{preskill}, can reveal a
connection between these two domains.

In statistical mechanics, the partition function captures how a
system’s microscopic configurations contribute to its macroscopic
thermodynamic properties in equilibrium. In contrast, the Loschmidt
amplitude plays a central role in quantum dynamics and information
theory, measuring the overlap between an evolving quantum state and
its initial configuration. Despite their different contexts, we show
that both quantities can be written as extensions of a single analytic
function in the complex plane.

To develop this connection, we begin with a single qubit, chosen as a
minimal prototypical system described by a finite-dimensional Hilbert
space, where the central concepts can be illustrated without much
technical overhead.  We observe that both the partition function and
the Loschmidt amplitude can be expressed as a polynomial in a complex
variable \( y \), whose roots encode unifying features of thermal and
dynamical behavior. While the model lacks true phase transitions, the
zeros still reflect important signatures such as bounds imposed by
quantum speed limits and logarithmic divergences in the rate
functions. The framework is then generalised to many-body systems,
including a chain of non-interacting qubits and an interacting spin
system. Throughout, we take a pedagogical approach, highlighting how
basic models can illuminate the analytic structure underlying both
equilibrium and nonequilibrium quantum physics.
% \vspace{-1em}
\subsection{Outline}
This paper is organized as follows. \hyperref[sec:setup]{Section~II}
introduces the central model and describes the physical setup. While
\hyperref[TD:QSM]{Section~III} discusses the thermal description of
the qubit using the canonical partition function,
\hyperref[QD:UTE]{Section~IV} turns to the quantum dynamics of the
same system at zero temperature. We compute the Loschmidt amplitude,
and identify special times when the system becomes orthogonal to its
initial state.  \hyperref[AUF:PFLA]{Section~V} develops an analytic
framework that brings together thermal and quantum descriptions and
\hyperref[ETMBS]{Section~VI} extends this framework to many-body
settings. Starting with a chain of \( N \) non-interacting qubits, we
move towards interactions by considering an interacting spin chain
with open boundary conditions. \hyperref[C]{Section~VII} concludes
with reflections on the significance of the unifying analytic
framework as well as its pedagogical potential.
\hyperref[Appendix]{Appendix~A} discusses the
Mandelstam--Tamm and Margolus--Levitin bounds in the context of our
specific models. Appendix \ref{sec:Zeno_appendix} discusses the
quantum Zeno paradox that is based on the early time behaviour
developed in Sec. \ref{HTLETE}.  \hyperref[AppendixB]{Appendix~C}
outlines a project idea and a set of homework exercises designed to
reinforce the concepts developed in the main text.

\section{The Two-Level System: Setup} \label{sec:setup}

\begin{figure}[htbp]
    \centering
    \begin{minipage}{0.9\linewidth}
        \centering
        \begin{minipage}{0.48\linewidth}
            \centering
            \includegraphics[width=\linewidth]{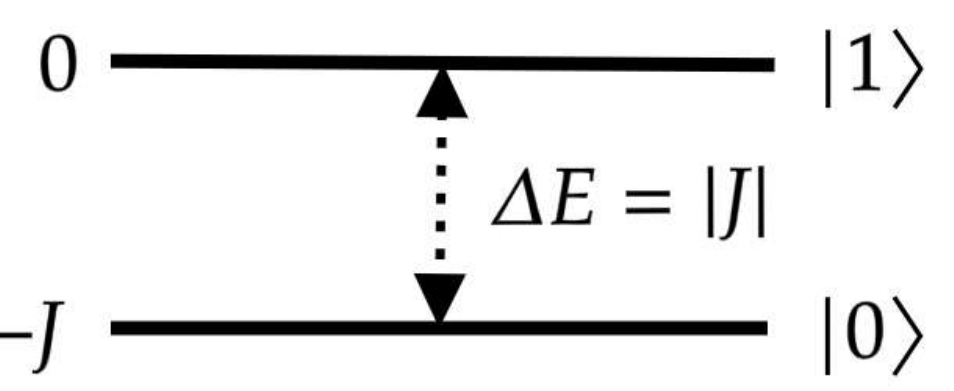}
        \end{minipage}
        \hfill
        \begin{minipage}{0.48\linewidth}
            %  % Adjust vertical alignment
            \begin{align*}
                \langle 0 | 0 \rangle = \langle 1 | 1 \rangle &= 1 \\
                \langle 0 | 1 \rangle = \langle 1 | 0 \rangle &= 0
            \end{align*}
        \end{minipage}
        \caption{A schematic diagram of a qubit's energy levels with
          an energy gap $\Delta E$ of magnitude $J (>0)$ between the
          two levels. On the right, the orthonormality conditions of
          the basis states are shown.} 
        \label{fig:two_level}
    \end{minipage}
\end{figure}

Let us consider a qubit with $\Omega=2$ non-degenerate energy
levels. The Hamiltonian lives in a two-dimensional Hilbert space, with
the energy spectrum given by
\begin{equation}
H|0\rangle = -J|0\rangle, \quad H|1\rangle = 0|1\rangle
\label{eq:Hamiltonian}
\end{equation}
where $|0\rangle$ and $|1\rangle$ form an orthonormal basis of energy
eigenstates, see Fig. \ref{fig:two_level}. This defines the simplest
non-trivial quantum system. In quantum information, this system
naturally serves as a qubit, the basic unit of quantum computation.
Another physical realization of a qubit is a spin-$\frac{1}{2}$ particle in a
magnetic field, where the energy levels correspond to spin-up and
spin-down along the field axis.

\section{Thermal Description: Quantum Statistical Mechanics} \label{TD:QSM}

The thermodynamic properties of the qubit, with $\Omega=2$ states and
coupled to a heat bath at inverse temperature \( \beta = 1/(k_B T) \),
where \( T \) is the temperature and \( k_B \) is the Boltzmann
constant, are governed by the partition function~\cite{pathria},
\begin{equation}
  Z(y) = \frac{1}{\Omega}\sum_{\alpha = 0}^{1} e^{-\beta E_\alpha} = \frac{1}{2}(1 + y), \quad y = e^{\beta J},
    \label{eq:partition_function}
\end{equation}
with \( E_0 = -J \) and \( E_1 = 0 \) as defined in
Eq.~\eqref{eq:Hamiltonian}.  The prefactor $1/\Omega$ simply removes
the contribution of the high-temperature entropy, $k_B \ln \Omega$,
from the free energy.  For the case of interest, $J>0$, the parameter
$y$ takes values in $[1,\infty)$ \cite{pathria}.  A useful feature is
that the partition function $Z(y)$ is analytic and strictly positive
for all real $\beta$.
    
\section{Quantum Description: Unitary Time Evolution} \label{QD:UTE}

In contrast to the thermal description, where the qubit is in contact
with a heat bath, we now examine the dynamics of the same qubit, but
in isolation and at zero temperature. With the energy conserved, its
time evolution is governed by the unitary operator
$U(t) = e^{-iHt/\hbar}$ ($\hbar$ being the reduced Planck constant).

Initially, we prepare the qubit in the superposition state
\begin{equation}
    |\psi(0)\rangle = \frac{1}{\sqrt{\Omega}}(|0\rangle + |1\rangle),\quad (\Omega=2),
    \label{eq:initial_state}
\end{equation}
which is clearly not an energy eigenstate of the Hamiltonian, and it evolves in time as
\begin{equation}
    |\psi(t)\rangle = e^{-iHt/ \hbar}|\psi(0)\rangle = \frac{1}{\sqrt{2}}(y|0\rangle + |1\rangle), \quad y = e^{iJt/\hbar}.
    \label{eq:state_evolution}
\end{equation}

\begin{figure}[htbp]
    \centering
    \includegraphics[width=\linewidth]{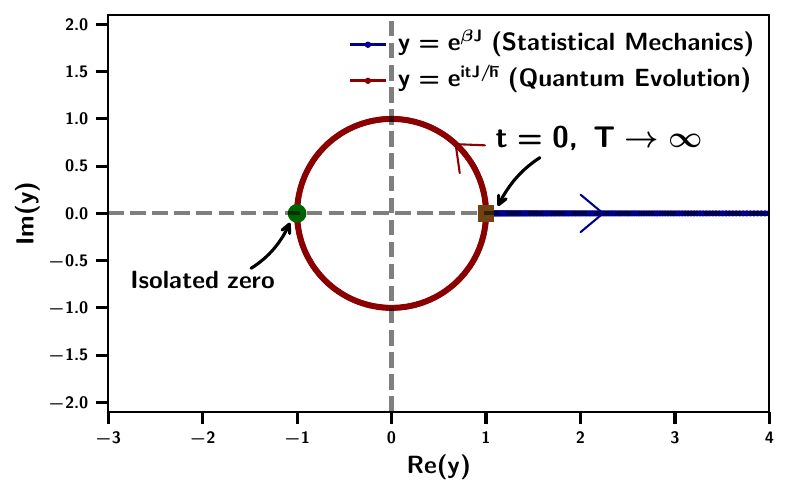}
    \caption{The blue line represents the real Boltzmann factor from
      statistical mechanics, which never intersects the isolated zero
      at \( y = -1 \) (the green point). The red unit circle
      corresponds to quantum time evolution.  It traces the complex
      Boltzmann factor, periodically crossing the zero.  The brown
      point at \( y = 1 \) marks both the infinite-temperature limit
      \(\beta \to 0\) and the initial time of the quantum system \(t
      \to 0\).}
    \label{fig:periodicity}
\end{figure}

Here \( y = e^{iJt/\hbar} \) is the natural analytic continuation of the
variable defined in Eq.~\eqref{eq:partition_function}. This
\textit{complex Boltzmann factor} traverses a unit circle as $t \in
(-\infty, \infty)$ \cite{negtime} in the complex-$y$ plane, see Fig.
\ref{fig:periodicity}. The unitary evolution introduces a relative
phase between the energy eigenstates, though the individual state
probabilities remain constant,
\begin{equation}
   P_{\alpha}(t) = |\langle \alpha|\psi(t)\rangle|^2 = \frac{1}{2}, \quad \alpha = 0, 1.
    \label{eq:P0_P1}
\end{equation}
For the dynamical behavior, we examine the Loschmidt amplitude, the
overlap between the initial and time-evolved states
\begin{equation}
    L(y) = \langle \psi(0) | \psi(t) \rangle = \frac{1}{2} \sum_\alpha e^{-iE_\alpha t/\hbar}= \frac{1}{2}(1 + y).
    \label{eq:loschmidt_amplitude}
\end{equation}

This overlap determines how distinguishable the time-evolved state is
from the initial one; when \( L(y) = 0 \), the states are orthogonal
and perfectly distinguishable.

The return probability (also known as the \textit{Loschmidt echo})
provides the likelihood of the system to return to its initial state
after a time $t$ and is given by,
\begin{equation}
    P(y) = |L(y)|^2=\frac{1}{2} [1 + \cos(Jt)]= \cos^2(Jt/2), 
    \label{eq:return_probability}
\end{equation}
where we used $y=\cos(Jt) + i \sin(Jt)$.

This quantity oscillates in time, and the qubit returns to its initial
state periodically. These
oscillations are analogous to Rabi cycles seen in two-level systems
under resonant driving, though here they arise purely from free
evolution. The return probability vanishes when $\cos(Jt) = -1$, at
\begin{equation}
  t_{c,n}  = \frac{(2n + 1)\pi \hbar}{J}, \quad n \in \mathbb{Z}.
    \label{eq:critical_times_1}
\end{equation}
At these special times, the system reaches a state orthogonal to the
initial one, a physically distinguishable configuration. The minimum
time required to reach an orthogonal state (in proper units) is thus
given by
\begin{equation}
    t_c = \frac{\pi \hbar}{J}, \quad (n=0).
    \label{eq:critical_times}
\end{equation}
The quantum speed bounds proposed by Mandelstam–Tamm and
Margolus–Levitin are saturated in this
case~\cite{frey,deffner2017qsl}. For a discussion of these limits in
the context of a qubit, see appendix \ref{Appendix}.

\section{A Unified Framework: Partition Function and Loschmidt Amplitude} \label{AUF:PFLA}
A connection between thermal and quantum descriptions of the qubit arises when we observe that both the partition function (Eq.~\ref{eq:partition_function}) and the Loschmidt amplitude (Eq.~\ref{eq:loschmidt_amplitude}) share a common algebraic structure. We define a dynamical partiton function \( \mathcal{L}(y) \) via analytic continuation, using \( y \) as a complex variable,
\begin{equation}
\mathcal{L}(y) = \frac{1}{2}(1 + y) =
\begin{cases}
Z(y), & \quad y = e^{\beta J}, \quad y \in \mathbb{R}^+ \\
L(y), & \quad y = e^{iJt/\hbar}, \quad |y| = 1.
\end{cases}
\label{eq:L_cases_combined}
\end{equation}

This unified polynomial structure allows us to interpret the Loschmidt
amplitude and the partition function as two manifestations of a single
analytic object, differing only in the path traversed in the
complex-\( y \) plane. Just as the partition function yields the
(dimensionless) thermal free energy,
\begin{equation}
f_{\text{thermal}}(y) \sim -\ln Z(y),
\label{eq:f_thermal}
\end{equation}
we define a free energy-like quantity, which we refer to as the
dynamical free energy,
\begin{equation}
f_{\text{dyn}}(y) \sim -\ln {L}(y).
\label{eq:f_qd}
\end{equation}
It is a complex-valued function, with its real part
\begin{equation}
f_L(y) = \text{Re}(f_{\text{dyn}}(y))=-\ln |L(y)|,
\label{eq:real_dyn}
\end{equation}
determining the return probability via
\begin{equation}
P(y) = \exp\left[-2 f_L(y)\right].
\label{eq:P_y_exp_f}
\end{equation}
The key parallels between thermodynamics and quantum dynamics for a
qubit are summarized in Table \ref{tab:thermal_quantum_comparison}.

\noindent 
\begin{table}[!b]
\centering
\renewcommand{\arraystretch}{1.6}
\setlength{\tabcolsep}{6pt}
\footnotesize
\rowcolors{2}{gray!10}{white}
\begin{tabularx}{\linewidth}{|Y|Y|Y|}
\hline
\rowcolor{gray!25}
\textbf{Quantity} & \textbf{Thermal Case} & \textbf{Quantum Case} \\
\hline
Variable & \( y = e^{\beta J} \in \mathbb{R}^+ \) & \( y = e^{iJt/\hbar},\ |y| = 1 \) \\
Function & \( Z(y) = \frac{1}{2}(1 + y) \) & \( L(y) = \frac{1}{2}(1 + y) \) \\
Free energy & \( f_{\text{thermal}}(y) \sim -\ln Z(y) \) & \( f_{\text{dyn}}(y) \sim -\ln L(y) \) \\
Free energy behavior & Continuous and finite & May exhibit logarithmic divergences \\
Physical observable & \( Z(y) \sim e^{-f_{\text{thermal}}} \) & \( P(y) \sim e^{-2 f_L(y)} \) \\
Energy fluctuation near \( y = 1 \) 
& \( -\beta^2 \dfrac{d^2 f_{\text{thermal}}}{d\beta^2} \Big|_{y=1} \) 
& \( t^2 \dfrac{d^2 f_L}{dt^2} \Big|_{y=1} \) \\
Interpretation & High-temperature specific heat & Early-time unitary evolution \\
\hline
\end{tabularx}
\caption{Comparison of thermal and quantum formulations in a two-level
  system, highlighting the role of the analytic function \(
  \mathcal{L}(y) \) and its implications. The correspondence between
  fluctuation terms links high-temperature thermodynamics and
  short-time quantum dynamics near \( y = 1 \).} 
\label{tab:thermal_quantum_comparison}
\end{table}

\subsection{High-Temperature Limit and Early-Time Expansion:
  Cauchy-Riemann Equations} \label{HTLETE} 

We now analyze the behavior of \(\mathcal{L}(y)\) in the vicinity of
\( y = 1 \), which corresponds to the infinite-temperature limit in
the thermal case and the early-time limit in the quantum evolution. In the
thermal regime, the second derivative of the free energy determines
the specific heat,
\begin{equation}
    C(\beta) = -k_B\, \beta^2 \frac{d^2 f_{\text{thermal}}}{d\beta^2}.
    \label{eq:thermal_sp_heat}
\end{equation}
This gives specific heat at high-temperatures as, 
\begin{equation}
    C(\beta) \,\,\,\,\, \underset{\mathclap{\text{  high } T}}{\approx} \,\,\,\, \frac{k_B}{4}\,J^2 \beta^2.
    \label{eq:thermal_sp_heat_small_beta}
\end{equation}
The $\beta^2$ dependence of specific heat in the high temperature
regime is generic for systems with a finite dimensional Hilbert space.

Since \( f_{\text{dyn}}(y) \) is analytic in a neighborhood of \( y =
1 \), see Eq. (\ref{eq:L_cases_combined}) and Eq. (\ref{eq:f_qd}), the
Cauchy-Riemann conditions apply and allow us to expand the dynamical
free energy \( f_{\text{dyn}}(y) \) in a Taylor series around \( y = 1
\). These conditions guarantee that the complex derivatives are
independent of the directions along which the point is approached. We
choose two orthogonal directions: (i) along the real axis and (ii)
along the tangent to the unit circle \(|y|=1\).  See Fig. \ref{fig:periodicity}. These correspond,
respectively, to high-temperature expansion in \( \beta \) and
early-time expansion in \( t \).  As a
result, just as the high-temperature specific heat reflects energy
fluctuations which also appears in the coefficient of the
\( t^2 \) term in the Taylor expansion of \( f_L(y)\). 

This correspondence links
the high-temperature thermodynamic behavior
(under $\beta\leftrightarrow t$)  to early-time evolution in the
quantum regime, see 
Table
\ref{tab:thermal_quantum_comparison}. In particular, for early times
i.e. \( \frac{Jt}{\hbar} \ll 1 \), this correspondence gives us
\begin{equation}
  f_L(t) \approx \frac{1}{2} 
  \left.\frac{C(\beta)}{k_B} \right|_{\beta \to t/\hbar} \,\,\,\,\,\underset{\mathclap{\text{small } t}}{\approx} \,\,\,\,\,\,\frac{1}{8 \hbar^2}J^2t^2, 
  \label{eq:CL_approx}
\end{equation}
where the second approximation follows from
Eq.(\ref{eq:thermal_sp_heat}) and
Eq.(\ref{eq:thermal_sp_heat_small_beta}).  What is striking is that a
high-temperature calorimetric quantity, namely, the specific heat,
maps onto the zero-temperature quantum dynamics of the system.

\subsubsection{Digression: Quantum Zeno effect}

The early-time quadratic dependence of $f_L(t)$ is central to
understanding the quantum Zeno effect \cite{misra,itano,itano2}. From
Eq. \eqref{eq:CL_approx} we see that the probability of
\textit{leaving} the initial state grows only quadratically at short
times. Independently, any measurement that checks for survival
collapses the evolved state onto the initial state. Repeated
alternation of short-time evolution with measurement-induced collapse
can substantially inhibit the system's overall evolution. Appendix
\ref{sec:Zeno_appendix} provides further discussion.

 \subsection{Zeros and Divergences} \label{ZAD}

\begin{figure}[htbp]
    \centering
    \includegraphics[width=\linewidth]{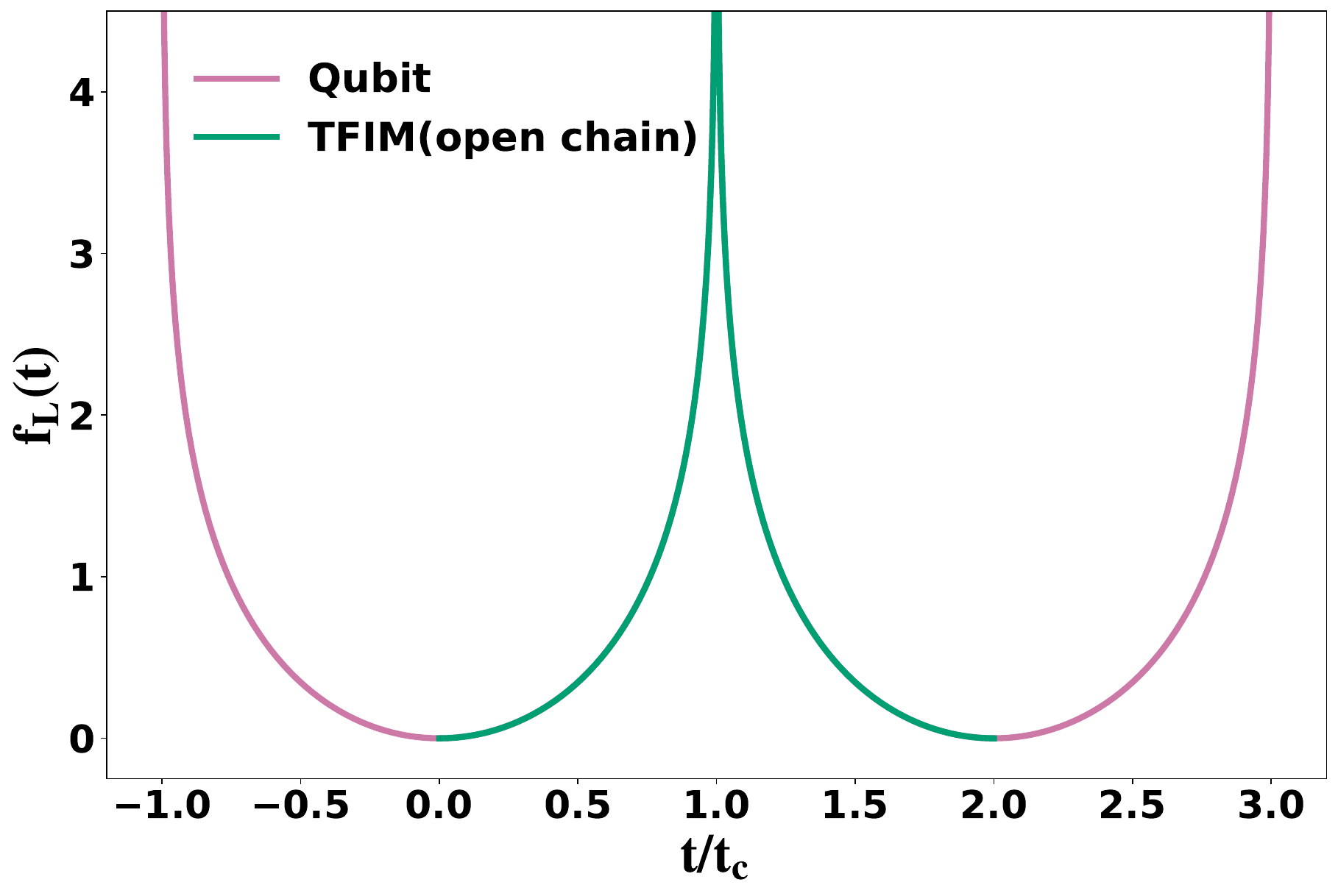}
    \caption{Periodic logarithmic divergences are observed in the real
      dynamical free energy \( f_L(y) \) as a function of scaled time
      in (i) a qubit (pink) and (ii) TFIM with open boundary conditions (green).
      }
    \label{fig:log divergences}
\end{figure}

For the qubit, \( \mathcal{L}(y)\) has a single isolated zero at
\begin{equation}
    y_0 = -1,
\end{equation}
as seen from Eq.~\eqref{eq:L_cases_combined}.  The importance of zeros
of the partition function was originally explored in the context of
phase transitions by Lee, Yang, and Fisher~\citep{pathria}, where the
limit points of complex zeros on the real axis signal singularities in
the thermodynamic limit. In general, the behavior of a polynomial is
governed by the location of its zeros.  In the thermal case, where \(
y = e^{\beta E} \in \mathbb{R}^+ \), the zero at \( y = -1 \) lies
outside the physical domain and remains inaccessible. In contrast, in
the quantum case, whenever the evolution path crosses the zero at \( y
= -1 \), \( f_L(y) \) diverges logarithmically, as shown in Fig.
\ref{fig:periodicity} and Fig.\ref{fig:log divergences}. This
indicates that in unitary dynamics, the system becomes orthogonal to
its initial state, causing the return probability to vanish and the
real dynamical free energy to diverge.

The possibility of crossing zero depends on the system’s energy
spectrum. For example, in a degenerate two-level system, such as when
the excited state \(\ E_1 =0\) as defined in
Eq.(\ref{eq:Hamiltonian}), is two-fold degenerate; the zero lies off
the unit circle, and the return probability remains nonzero at all
times. That is, the system never evolves into a state orthogonal to
its initial configuration. (See Appendix \ref{AppendixB}.)

\smallskip
\noindent
\textit{A side note.}  This framework also invites an electrostatic
analogy, that is especially natural in two dimensions. The zero of \(
\mathcal{L}(y) \) at \(y=-1\) can be interpreted as the location of a point
charge in the complex-\( y \) plane, with the dynamical free energy \(
f_{dyn}(y) \) playing the role of the electrostatic potential generated
by this charge \cite{leeyang,fisher2}. For the simple qubit system, this
analogy is mainly illustrative, but in more complex many-body systems,
it becomes a powerful way to understand how phase transitions arise \cite{heylreview}.
The qubit example nevertheless conveys the essential idea, that the
zeros in the complex $y$-plane act like charges whose positions
determine both the statistical mechanical behaviour and the
zero-temperature quantum evolution.

\subsection{On initial state}
\label{sec:init-state}

The particular initial state of  Eq. \eqref{eq:initial_state} is motivated by the high temperature limit where the two states occur with equal probability.  It is also useful to consider a more general normalized state,
\begin{equation}
    |\psi(0)\rangle = \frac{1}{\sqrt{|c_0|^2+ |c_1|^2}}(c_0|0\rangle + c_1 |1\rangle),
   \label{eq:initial_state2}
\end{equation}
where $c_0,c_1$ are arbitrary complex amplitudes.

Following the same steps as before, the Loschmidt amplitude takes the form
\begin{equation}
\label{eq:g0g1}
 L(y)= g_1+ g_0 \,y, \quad {\textrm{ with }} g_0=\frac{|c_0|^2}{|c_0|^2+|c_1|^2}, 
\end{equation}
and $g_1=1-g_0.$   For real $y\geq 1$ and $0<g_0<1$, it is possible to interpret this $L(y)$ as a partition function with  nonthermal statistical weights for the states.  Note that we recover  Eq. \eqref{eq:loschmidt_amplitude} for $g_0=1/2$, whereas  the degenerate two level system considered in the previous subsection corresponds to $g_0=1/3$.

The zero of $L(y)$ is 
\begin{equation}
    y_0=-\frac{1-g_0}{g_0},
\end{equation}
which lies on the unit circle only when $g_0=1/2$.  Thus, for a generic initial state, the Loschmidt amplitude never vanishes; instead, the return probability exhibits a finite minimum at the time $t=tc$ given by Eqs.~\eqref{eq:critical_times_1} and \eqref{eq:critical_times}. The degree of “non-orthogonalizability’’ can be quantified by the maximum of
$f_L(y)$
which occurs at $t=t_c$. 
Denoting the minimum distance of the zero, $y_0$, from the unit circle by 
$$\Delta = |y_0 +1|= \frac{|2g_0-1|}{g_0},$$
we find  that the maximum value  $f_{\textrm max}$  depends on $\Delta$ as 
\begin{equation}
    \label{eq:f-delta}
    f_{\textrm max} \sim -\ln \Delta, {\textrm{ as }} \Delta\to 0.
\end{equation}
The high-temperature specific heat continues to exhibit a $\beta^2$ dependence, with a coefficient that depends on $g_0$. Importantly, the correspondence between the specific heat and the early-time behavior of the quantum dynamics remains unchanged.

Even in this more general setting, the thermal behavior may still be viewed through the electrostatic analogy: $f(y)$ corresponds to the two-dimensional potential generated by a point charge located at $y=y_0.$\\[6pt]

\section{Extension to Many-Body Systems} \label{ETMBS} 

We now examine
the collective behavior of a many-body system under two complementary
approaches: thermal equilibrium (when the system is connected to a
heat bath) and unitary quantum dynamics (when the system is isolated).
To this end, we consider a chain of \( N \) non-interacting qubits,
living in a \(2^N\)-dimensional Hilbert space.

In the thermal setting, the additivity of the Hamiltonian implies that
the total partition function factorizes as
\begin{equation}
Z_N(y) = [Z(y)]^N = \left( \frac{1 + y}{2} \right)^N,
\label{eq:Z_N}
\end{equation}
where \( Z(y) \) is the single-qubit partition function defined
earlier in Eq.~\eqref{eq:partition_function}, and \( Z_N \) is the
partition function of the full \( N \)-qubit chain. The free energy
per qubit in the thermodynamic limit is then given by
\begin{equation}
f_{\text{thermal}}(y) = \lim_{N \to \infty} -\frac{1}{N} \ln Z_N(y) = -\ln\left( \frac{1 + y}{2} \right).
\label{eq:free_energy_density}
\end{equation}

On the quantum side, we consider an initial product state of the form  
\begin{equation}
\ket{\Psi(0)} = \prod_{j=1}^{N} \ket{\psi_j(0)},
\label{eq:initial_state_1}
\end{equation}
where each \( \ket{\psi_j(0)} \) is the initial state of qubit \( j \)
as defined in Eq.(\ref{eq:initial_state}). Under unitary evolution
with a non-interacting Hamiltonian, the state at time \( t \) becomes
\begin{equation}
\ket{\Psi(t)} = \prod_{j=1}^{N} \big( e^{-i H t/ \hbar} \ket{\psi_j(0)}\big).
\end{equation}

The corresponding Loschmidt amplitude for the many-body system is the
product of single-qubit amplitudes, allowing us to write
\begin{equation}
\mathcal{L}_N(y) = [\mathcal{L}(y)]^N,
\end{equation}
where \( \mathcal{L}(y) \) is the dynamical partition function associated with a single
qubit, as defined in Eq.~\eqref{eq:L_cases_combined} and \(
\mathcal{L}_N(y) \) denotes the  corresponding quantity for the full
system. Following the single-qubit case, we define a quantum analog of
the free energy density per qubit in the thermodynamic limit:
\begin{equation}
f_{\text{dyn}}(y) \equiv \lim_{N \to \infty} -\frac{1}{N} \ln \mathcal{L}_N(y) = -\ln \mathcal{L}(y).
\label{eq:f_qd_density}
\end{equation}
The return probability in the large-\( N \) limit becomes  
\begin{equation}
P(y) = \exp\left[-2N f_L(y)\right],
\label{eq:return_prob_large_N}
\end{equation}
where $f_L(y)$ is given by Eq.\eqref{eq:real_dyn}.
The exponential form of Eq.~\eqref{eq:return_prob_large_N} plays a
central role in large deviation theory~\cite{touchette}, where the
probabilities of rare fluctuations decay exponentially with system
size.

\subsection{Toward Interactions} \label{TI} We now examine how the
analytic structure discussed above extends to systems with
interactions. Specifically, we consider a one-dimensional chain of \(
N \) spin-\( \frac{1}{2} \) particles with open boundary conditions,
defined in a \( 2^N \)-dimensional Hilbert space.

Each site \( i \) hosts a spin described by the Pauli operators \(
\sigma_i^x, \sigma_i^y, \sigma_i^z, \) which are defined as
\[
\sigma^x = \begin{pmatrix}
0 & 1 \\
1 & 0
\end{pmatrix}, \quad
\sigma^y = \begin{pmatrix}
0 & -i \\
i & 0
\end{pmatrix}, \quad
\sigma^z = \begin{pmatrix}
1 & 0 \\
0 & -1
\end{pmatrix}.
\]
In this section, we focus on an
interaction Hamiltonian involving only the \( \sigma^z \) operators:
\begin{equation}
H = -\frac{J}{2} \sum_{i=1}^{N-1} \sigma_i^z \sigma_{i+1}^z,
\label{eq:ising_H}
\end{equation}
where \( J/2 \) sets the nearest-neighbor interaction strength.

This Hamiltonian is diagonal in the basis of product eigenstates of
the \( \sigma_i^z \) operators. Since each spin can take eigenvalues
\( \pm 1 \), the Hilbert space consists of \( 2^N \) distinct
configurations of the form \( \{s_1, s_2, \dots, s_N\} \), where \(
s_i = \pm 1 \) is the eigenvalue of \( \sigma_i^z \) at site \( i \).
The states \( |\uparrow\rangle \) and \( |\downarrow\rangle \) refer
to the eigenstates of \( \sigma^z \) with eigenvalues \( +1 \) and \(
-1 \), respectively. A configuration is thus a sequence of \( N \)
such eigenstates. The corresponding energy eigenvalue of the chain is
\begin{equation}
E = -\frac{J}{2} \sum_{i=1}^{N-1} s_i s_{i+1},
\end{equation}
which is precisely the \textit{classical} Ising model.

In the thermal setting, the partition function is
\begin{equation}
Z_N = \text{Tr} \,e^{-\beta H} = \sum_{\{s_i = \pm 1\}} e^{\beta \frac{J}{2} \sum_{i=1}^{N-1} s_i s_{i+1}}.
\end{equation}
To simplify this sum, we introduce bond variables \( \zeta_i = s_i
s_{i+1} \in \{\pm 1\} \), so that the energy becomes
\begin{equation}
E = -\frac{J}{2} \sum_{i=1}^{N-1} \zeta_i.
\end{equation}
To simplify the spectrum, we shift this energy expression by a constant, 
which does not affect the physics, as
\begin{equation}
E = -J\sum_{i=1}^{N-1} \left(\frac{1+\zeta_i }{2} \right),
\label{eq:shifted_Ising}
\end{equation}
so that the two energy levels per bond are illustrated in
Fig.~\ref{fig:two_level}. The minimal excitation corresponds to
flipping a bond, giving an energy gap \(\Delta E = J\). Because the
bond variables are independent, the partition function factorizes
across bonds. This leads to \cite{qubitbond}

\begin{equation}
Z_N(y) = \left(\frac{1+y}{2}\right)^{N-1}.
\end{equation}
This reproduces the same polynomial form as in the non-interacting
case with \(N-1\) such zeros lying at \( y = -1 \), still being
outside the thermal domain.

We now turn to the quantum dynamics of the same model. The system is
initially prepared in a product state that is not an eigenstate of the
Hamiltonian. One such choice is the ground state of a strong
transverse field, which aligns all spins along the \( \hat{x}
\)-direction. At \( t = 0 \), the transverse field is quenched to
zero, and the system evolves under the Ising interaction described in
Eq.~\eqref{eq:ising_H}. The corresponding initial state
\begin{equation}
\ket{\Psi(0)} = \prod_{j=1}^{N} \ket{\psi_j(0)}, \quad \ket{\psi_j(0)} = \frac{1}{\sqrt{2}} \left( |\uparrow_j\rangle + |\downarrow_j\rangle \right),
\end{equation}
is not an eigenstate of the Hamiltonian, ensuring non-trivial dynamics.

The Loschmidt amplitude can be computed as the overlap between the
initial state and the time-evolved state. Using
Eq.(\ref{eq:shifted_Ising}), this leads to
\begin{equation}
\mathcal{L}_N(y) = \frac{1}{2^{N}} \sum_{\{s_j = \pm 1\}} e^{-i E t/ \hbar} = \left(\frac{1+y}{2}\right)^{N-1}.
\label{eq:TFIM_zeros}
\end{equation}
This expression shares the same analytic structure as the thermal
partition function and exhibits a zero at \( y = -1 \). Thus, under
certain interactions and boundary conditions, the rate function
governing the collective behavior of the system retains the same
features as observed in the single-qubit case.

The effect of interactions becomes visible in the periodic-chain case, 
which we suggest as a project in Appendix~\ref{AppendixB}. For an interacting 
system, if the zeros of the partition function pinch the positive real axis in the infinite-size limit, a thermal phase transition occurs \cite{pathria,leeyang,fisher2}. Likewise, in the complex-$y$ plane, if the zeros in the infinite-size limit pinch or intersect the unit circle, the quantum evolution exhibits nonanalytic behavior in time. These singularities are examples of dynamical quantum phase transitions \cite{heylreview,heylprl,zvyagin,wei,peng}. 

If the zeros intersect the $|y|=1$ unit circle at isolated points, the
corresponding divergences in $f_L$ signal the orthogonalization of the
states at those times.  In other cases, the nature of the dynamical
quantum phase transitions will be determined by the density of zeros.

\section{Pedagogic values}
\label{sec:pedagogic-values}

\subsection{For students and instructors}
\label{sec:students-instructors}

A unified understanding becomes possible because both statistical
mechanics and quantum dynamics fundamentally rely on combining
contributions from all accessible states; it is through summation over
states, justified by ergodicity, in the former and through coherent
superposition in the latter. Once this common foundation is
recognized, it becomes evident that the familiar equilibrium and
time-dependent descriptions arise as different ways of exploring a
single underlying complex function.

For students, this perspective offers a concrete and computable
example in which the complex plane and analytic continuation become
physically meaningful, linking equilibrium and dynamical quantities
through different paths in the complex plane. They also see how ideas
from complex analysis, statistical mechanics, and quantum dynamics
come together within one accessible framework.

For teachers, it provides a compact, analytically transparent
framework that can be integrated into courses to bridge statistical
mechanics and quantum mechanics without heavy formalism, strengthening
coherence across the curriculum.

\subsection{On analytic continuation}
\label{sec:analyt-cont}

The observation that the Boltzmann factor in statistical mechanics and
the unitary time evolution operator in quantum mechanics are related
by analytic continuation is exploited in the path integral formulation
of a many-body system \cite{mahan}. In such a formulation, quantum
dynamics is expressed in terms of trajectories in space-time, and
thermal equilibrium is obtained by replacing real time with imaginary
time ($t \to i t$).  This time-transformtion, called Wick rotation,
converts a quantum problem with a Hamiltonian in $d$-dimensions into a
classical statistical system living in $(d+1)$-dimensios.  For
example, the two-state problem would become a one-dimensional spin
chain. See, e.g, Ref. \cite{mahan}.

Our approach takes a different path. By working directly with the
Loschmidt amplitude, which plays a role analogous to the diagonal
Green's function in the many-body framework of Ref.~\cite{mahan}, we
sidestep the path-integral formalism entirely. This means that neither
Wick rotation nor any additional mathematical apparatus need to be
introduced. The method relies solely on concepts students typically
encounter in standard courses in mathematical physics, statistical
mechanics, and quantum mechanics. In this way, it offers an accessible
means of understanding how equilibrium and dynamical descriptions are
connected, without requiring tools beyond the undergraduate
curriculum. Moreover, the same framework extends naturally to
research-level topics, such as dynamical quantum phase transitions, in
which the zeros of the Loschmidt echo in the complex-$y$ plane govern
the behavior of the transition \cite{heylreview,heylprl,zvyagin}.  The
problem of the Hamiltonian of Sec. \ref{TI} under periodic boundary
conditions can be a project topic for advanced courses in statistical
mechanics or quantum mechanics.\\[6pt]

\section{Conclusion} \label{C}

\noindent The central idea unifying the thermal and dynamical
descriptions is that both the partition function and the Loschmidt
amplitude can be expressed as evaluations of a single analytic
function \( \mathcal{L}(y) \) along different paths in the complex
\( y \)-plane.  With $-J$ and $0$ as the two energy values,
\( y = e^{\beta J} \) lies on the positive real axis in thermal
physics, while in quantum dynamics, \( y = e^{i J t/\hbar} \) traces
the unit circle. This analytic continuation shows that features such
as orthogonality and singularities in dynamical free energy are
governed by the same underlying zero structure that also encodes
thermal properties.  Throughout this work, the discussion is
restricted to examples defined on finite-dimensional Hilbert spaces,
where the spectrum is purely discrete, allowing this correspondence to
be demonstrated transparently using elementary tools.

The qubit-based approach presented here is both accessible and
rigorous, and thus can easily be incorporated in a statistical
mechanics or a quantum mechanics course. This framework opens a
pedagogical entry point for introducing advanced concepts such as
dynamical quantum phase transitions, and more
complex phenomena in interacting systems using only familiar tools
from undergraduate physics. Such advanced topics can further be
assigned as projects to the students, see
\hyperref[AppendixB]{Appendix~C}. As the connection between
equilibrium and nonequilibrium physics continues to unfold, the
deceptively simple qubit serves as a reminder that deep insights often
emerge from minimal models.

\vspace{1em}
\noindent\textbf{Conflict of interest} The authors declare no conflict of interest.\\

\appendix 
\section{Quantum Speed Limits (QSLs)} \label{Appendix} At a
fundamental level, quantum mechanics imposes a natural constraint on
how fast a qubit, or more generally any quantum system, can evolve in
time. In particular, there exists a minimum (\textit{first}) time
required for a system to evolve from an initial state to one that is
orthogonal to it. This \textit{orthogonalization time}, denoted by \(
\tau_\perp \), is defined by the condition \( \langle \psi_0 |
\psi(\tau_\perp) \rangle = 0.  \)

In our qubit model, this time is found to be
\begin{equation}
    t_c = \frac{\pi \hbar}{J},
    \label{eq:tau_perp}
\end{equation}
as discussed in Eq.~(\ref{eq:critical_times}) in the main text.

Quantum speed limits (QSLs) place lower bounds on \( \tau_\perp \).
These bounds arise from the spectral properties of the Hamiltonian and
the system’s energy distribution in the initial state. Two key
bounds~\cite{frey} are commonly used,

\begin{itemize}
    \item \textbf{Mandelstam--Tamm (MT) bound:}

    This bound is determined by the energy uncertainty \( \Delta E \), defined as
    \(
    \Delta E = \sqrt{\langle H^2 \rangle - \langle H \rangle^2}.
    \)
    It gives
    \begin{equation}
        \tau_\perp \geq \tau_{\text{MT}} = \frac{\pi \hbar}{2 \Delta E}.
        \label{eq:mt-bound}
    \end{equation}

    \item \textbf{Margolus--Levitin (ML) bound:}

    This bound uses the average energy above the ground
    state. Assuming the spectrum is shifted so that \( \varepsilon_n =
    E_n - E_0 \), the average shifted energy is 
    \(
    \langle \varepsilon \rangle = \langle H \rangle - E_0,
    \)
    and the bound takes the form
    \begin{equation}
        \tau_\perp \geq \tau_{\text{ML}} = \frac{\pi \hbar}{2 \langle \varepsilon \rangle}.
        \label{eq:ml-bound}
    \end{equation}
\end{itemize}

Since both bounds are valid and independent, the tightest constraint
on the orthogonalization time is given by
\begin{equation}
    \tau_\perp \geq \max\left( \tau_{\text{MT}}, \tau_{\text{ML}} \right) = \max\left( \frac{\pi \hbar}{2 \Delta E}, \frac{\pi \hbar}{2 \langle \varepsilon \rangle} \right).
    \label{eq:quantum-speed-limit}
\end{equation}

Both bounds vanish in the classical limit \( \hbar \to 0 \), as expected. We now derive each bound and evaluate them explicitly for the qubit.

\subsection{Mandelstam--Tamm Bound}

We begin with the uncertainty relation between two Hermitian operators
\( A \) and \( B \):
\begin{equation}
\Delta A \, \Delta B \geq \frac{1}{2} \left| \langle [A, B] \rangle \right|.
\label{eq:uncertainty_general}
\end{equation}
Taking \( B = H \) and using the Heisenberg equation \( [A, H] =
-i\hbar \frac{dA}{dt} \), we obtain
\begin{equation}
\Delta A \, \Delta H \geq \frac{\hbar}{2} \left| \frac{d}{dt} \langle A \rangle \right|.
\label{eq:uncertainty_applied}
\end{equation}

Choosing \( A = |\psi_0\rangle \langle \psi_0| \), the projection
operator onto the system's initial state, the return probability \(
P(t) = |\langle \psi_0 | \psi(t) \rangle|^2 \) corresponds to the
expectation value of \( A \) in the time-evolved state. The variance
of \( A \) is then given by \( P(1 - P) \). Substituting into
Eq.~\eqref{eq:uncertainty_applied} gives
\begin{equation}
\Delta H \geq \frac{\hbar}{2} \frac{1}{\sqrt{P(1 - P)}} \left| \frac{dP}{dt} \right|.
\label{eq:deltaH_bound}
\end{equation}

Integrating over time up to \( \tau_\perp \), where \( P(\tau_\perp) = 0 \), we find
\begin{equation}
\tau_\perp \geq \frac{\hbar}{2\Delta H} \int_1^{0} \frac{dP}{\sqrt{P(1 - P)}} = \frac{\pi \hbar}{2 \Delta H} = \tau_{\text{MT}}.
\label{eq:MT_bound}
\end{equation}

For the qubit initialized in the \textit{cat} state
(Eq.~\eqref{eq:initial_state}) and evolving under the Hamiltonian in
Eq.~\eqref{eq:Hamiltonian}, the energy variance is
\begin{equation}
(\Delta H)^2 = \langle \psi_0 | H^2 | \psi_0 \rangle - \langle \psi_0 | H | \psi_0 \rangle^2 = \frac{J^2}{4},
\label{eq:deltaH_value}
\end{equation}
which gives \( \Delta H = J/2 \), independent of time. Substituting into Eq.~\eqref{eq:MT_bound} provides
\begin{equation}
\tau_{\text{MT}} = \frac{\pi \hbar}{J},
\label{eq:MT_final}
\end{equation}
in agreement with the exact orthogonalization time \( \tau_\perp \) in Eq.~\eqref{eq:tau_perp}.

\subsection{Margolus--Levitin Bound}
Let \( H |n\rangle = E_n |n\rangle \), and consider the initial pure state
\begin{equation}
|\psi(0)\rangle = \sum_n c_n |n\rangle, \quad \sum_n |c_n|^2 = 1,
\label{eq:ML_initial_state}
\end{equation}
with shifted energies \( \varepsilon_n = E_n - E_0 \). Under time
evolution, the state becomes
\begin{equation}
|\psi(t)\rangle = \sum_n c_n e^{-i \varepsilon_n t / \hbar} |n\rangle.
\label{eq:ML_time_evolved}
\end{equation}

The Loschmidt amplitude is
\begin{equation}
L(t) = \langle \psi(0) | \psi(t) \rangle = \sum_n |c_n|^2 e^{-i \varepsilon_n t / \hbar}.
\label{eq:ML_loschmidt}
\end{equation}

Taking the real part and applying the inequality 
\begin{equation}
    \label{eq:cos-sin}
    \cos \theta \geq 1
- \frac{2}{\pi} \theta - \frac{2}{\pi} \sin \theta, {\textrm{ for  }} 0\leq \theta\leq\pi, 
\end{equation}
we obtain
\begin{equation}
\text{Re} \, L(t) \geq 1 - \frac{2t}{\pi \hbar} \langle \varepsilon \rangle - \frac{2}{\pi} \, \text{Im} \, L(t),
\label{eq:ML_bound}
\end{equation}
with
\begin{equation}
\langle \varepsilon \rangle = \sum_n |c_n|^2 \varepsilon_n = \langle H \rangle - E_0.
\label{eq:ML_average_energy}
\end{equation}

At \( t = \tau_\perp \), the amplitude vanishes and both its real and
imaginary parts are identically zero, implying
\begin{equation}
0 \geq 1 - \frac{2 \tau_\perp}{\pi \hbar} \langle \varepsilon \rangle,
\end{equation}
which leads to the ML bound
\begin{equation}
\tau_\perp \geq \frac{\pi \hbar}{2 \langle \varepsilon \rangle} = \tau_{\text{ML}}.
\label{eq:ML_final_bound}
\end{equation}

In our qubit model (see, Eqs.  \eqref{eq:Hamiltonian},
~\eqref{eq:initial_state} and \eqref{eq:ML_time_evolved}), we have 
\begin{equation}
\langle \varepsilon \rangle = \frac{1}{2} J,
\label{eq:ML_qubit_average}
\end{equation}
so that
\begin{equation}
\tau_{\text{ML}} = \frac{\pi \hbar}{J},
\label{eq:ML_qubit_final}
\end{equation}
which again coincides with \( t_c \), indicating that the
Margolus--Levitin bound is also saturated in this model. 

\subsubsection{Proof of inequality, Eq. \eqref{eq:cos-sin}}

To prove the inequality given in Eq. \eqref{eq:cos-sin},  
let us consider 
$$k(\theta)=\frac{1- \cos \theta}{\theta + \sin \theta}, $$
Since  
$$\frac{dk}{d\theta}= \frac{\theta \sin\theta}{\theta+\sin\theta}>0,{\textrm{ for  }}\theta\in[0,\pi], $$
$k(\theta)$ is monotonically increasing function at least in the range $\theta\in[0,\pi],$
and 
\begin{equation}
\label{eq:ktheta}
    k(\theta)=\left\{ \begin{array}{ll}
                    0& {\textrm{ for }}\theta =  0\\
                    2/\pi& {\textrm{ for }}\theta =  \pi    
                    \end{array}\right.
\end{equation}
Therefore $ k(\theta) \leq 2/\pi,$ or, $\cos\theta \geq 1 - \frac{2}{\pi} \theta - \frac{2}{\pi} \sin\theta.$

%%%%%%%%%%%%%%%%%%%%%%%%%%%%
\subsection{Scaling of QSLs in Many-Body Systems}

In Sec.~\ref{ETMBS}, we consider a non-interacting qubit chain and an
interacting spin chain with bond-variable representation. In both
cases, the total Hamiltonian is additive over \( N \) effective
degrees of freedom (qubits or bonds). Consequently, the average energy
\( \langle E \rangle \) scales linearly with \( N \), while the energy
uncertainty \( \Delta E \) grows as \( \sqrt{N} \) for independent
product states. Substituting into the QSL bounds in
Eqs.~\eqref{eq:mt-bound} and \eqref{eq:ml-bound}, we find that \(
\tau_{\text{ML}} \sim 1/N \) and \( \tau_{\text{MT}} \sim 1/\sqrt{N}
\). For $N\to\infty$, both the bounds vanish and therefore \textit{do
  not} provide much information about the QSLs.

%%%%%%%%%%%%%%%%%%%%%%%%%%

\section{Repeated Measurement and the Quantum Zeno Effect}
\label{sec:Zeno_appendix}

The survival probability is the probability that a system remains in the
state in which it was initially prepared.
We now determine the survival probability over a total time $\tau$
when the system is subjected to repeated measurements of the initial
state at intervals of duration $\tau/n$ \cite{misra,itano,itano2}. 

The survival probability $P(t)$ 
has the short-time form
\begin{equation}
  \label{eq:1}
P(t) \approx \exp(-\alpha t^{2}),
\end{equation}
as obtained from the early time behaviour given by
Eqs.~(\ref{eq:CL_approx}) and (\ref{eq:P_y_exp_f}).  This quadratic
dependence reflects the general quantum-mechanical result that
departures from the initial state occur only to second order in time.

According to the collapse postulate of quantum mechanics, each
measurement projects the state onto the initial state whenever the
system is found there, after which it again evolves for a time
$\tau/n$ according to the same probability law $P(t)$ of
Eq. \eqref{eq:1}.  Because each short-time evolution is independent,
the survival probability after $n$ such cycles is
\begin{equation}
  \label{eq:2}
P_n(\tau)
 = [P(\tau/n)]^{n}
 = \left[ e^{-\alpha\, \tau^2/n^2} \right]^{n}
 =e^{-\alpha \tau^2/n}.
  \end{equation}
In the limit of increasingly frequent measurements,
\begin{equation}
  \label{eq:3}
\lim_{n\to\infty} P_n(\tau)= 1,
\end{equation}
so the system remains in its initial state with probability $1$.
Thus, frequent measurements inhibit the system's evolution away from
its initial state.  This phenomenon is known as the \textit{quantum
  Zeno effect}.

\section{Exercises}\label{AppendixB}
A possible project topic:
\begin{quote}
  Consider the Ising model of Section~\ref{ETMBS}A with periodic
  boundary conditions. Explore the features of phase transitions.
\end{quote}
This project naturally leads to the recently developed notion of
dynamical quantum phase transitions (DQPT)~\cite{heylreview,heylprl,zvyagin,wei,peng}. 
In this example, the orthogonalization is preempted by a DQPT singularity in $f_L$ at a critical time $t_c$.  There is no $\tau_{\perp}$.

\vspace{1em}

 The following homework problems can be assigned:
    \begin{enumerate}

    \item A two-level system with a degenerate excited state (two-fold
      degeneracy). Analyze the time evolution of the return
      probability, and connect its periodicity to the location of the
      zero.
        
    \item Repeat the above analysis for a harmonic oscillator with the
      initial state prepared in an equal superposition of the first
      three energy eigenstates.  Note that there are more than one
      zero.  Generalize the result to an initial state with an
      arbitrary number of superposed energy levels.
      \end{enumerate}


\begin{thebibliography}{99}
\bibitem{negT} Upon including the negative temperatures, a regime associated with
population inversion in certain systems~\cite{pathria}, the domain of
\( y \) extends to the interval \( (0, \infty) \).  For $J<0$, the range of $y$ would be $0$ to $1$ for positive temperatures with $y=1$ as the $T\to\infty$ limit.
\end{thebibliography}
\end{document}